\documentclass[a4paper]{jpconf}
\usepackage{graphicx}
\usepackage{amsmath} 
\usepackage{cite}
\usepackage{hyperref}
\usepackage{amsmath,amssymb}
\usepackage{color}
\begin{document}

\title{On the null energy condition in scale
dependent frameworks with spherical symmetry}
\author{\'Angel Rinc\'on, Benjamin Koch}
\address{Physics Institute, Pontifical Catholic University of Chile, Av. Vicu\~na Mackenna 4860, Santiago, Chile}
\ead{\href{mailto:arrincon@uc.cl}{\nolinkurl{arrincon@uc.cl}}}

\begin{abstract}
The effect of a scale dependent newton coupling, which is a crucial ingredient of many quantum gravity theories, is investigated in this paper. For case of non-rotating black holes, and using the null energy condition, we show that Newtons scale dependent coupling can actually be obtained without solving the full quantum gap equations.
\end{abstract}

\section{Introduction}
Nowadays, different alternatives exist to obtain an unified theory of quantum gravity. Loop quantum gravity \cite{book:217893} as well as string theory \cite{Polchinski:1998rq,Polchinski:1998rr} are famous candidates to achieve this unification. Besides, a self-consistent theory of quantum gravity is still lacking, being studied assuming different points of view \cite{Wheeler:1957mu,Deser:1976eh,Rovelli:1997yv,Bombelli:1987aa,Ashtekar:2004vs,Sakharov:1967pk,Jacobson:1995ab,Verlinde:2010hp,Reuter:1996cp,Litim:2003vp,Horava:2009uw,Charmousis:2009tc,Ashtekar:1981sf,Penrose:1986ca,Connes:1996gi,Nicolini:2008aj,Gambini:2004vz}. 
Many quantum extensions of general relativity present a common feature: a scale dependence appearing at the level of the effective action of gravity. 
The classical couplings are not constants any more, instead, they depend on arbitrary scale $k$. 
The physics of a black hole strongly depends on the
scale setting e.g. the connection between $k$ and $r$.
\noindent For the case of spherically symmetric black holes, it can be expected that $k=k(r)$ and thus $\{G_k, \Lambda_k, e_k, \mathcal{O}_k\} \rightarrow \{G(r), \Lambda(r), e(r), \mathcal{O}(r)\}$ \cite{Koch:2014joa,Koch:2010nn}. Those $r$-dependent couplings can be obtained directly from the quantum gap equations, if one assumes certain energy conditions.

\noindent Recent papers have shown that the so-called ``Null Energy Condition'' (hereafter NEC) plays a crucial role in this context \cite{Koch:2016uso,Contreras:2013hua,Rincon:2017ypd}. Thus, in this conference proceedings we analyze the link between the so-called Schwarzschild ansatz and the NEC, in the light of scale dependent couplings. We show farther the important role played by the NEC in determining the gravitational running coupling $G(r)$. The present work is organized as follows: after this short introduction, we explain the main idea of this letter in Section \ref{Main_Idea}, whereas the technique and results are collected at Section \ref{Technique}. Finally, the conclusions are given at Section \ref{Conclusion}. 

\section{Main Idea}\label{Main_Idea}
The so-called Null Energy Condition is one of the usual energy conditions (dominant, weak, strong, and null) and, besides, the least restrictive of them. It helps to obtain suitable solutions of the Einstein field equation and it is usually applied to discern whether a solution has physical validity or not. We will explore this condition in the context of spherical symmetry where the line element reads
\begin{align}\label{line_element}
ds^2 &= -f(r)dt^2 + g(r)dr^2 + r^2 d\Omega,
\end{align}
which have to be determined by theorems. Here
$f(r)$ and $g(r)$ are arbitrary functions, and $d\Omega$ is the solid angle (which depends on the studied dimension). 
In many cases the task of solving the EOM's is simplified by the so called Schwarzschild ansatz $f(r) \cdot g(r) = 1$. As was explained by Jacobson \cite{Jacobson:2007tj}, the aforementioned ansatz can be seen as a consequence of NEC. This finding follows from the fact that the Ricci tensor is proportional to the metric in the $t-r$ subspace. In addition, and regarding the application of this condition, one should note that validity of NEC was guaranteed at Ref. \cite{Jacobson:2007tj}, and it applies in spherical symmetry with either Maxwell electrodynamics (i.e. the Reissner-Nordstrom solution) or Born-Infeld non-linear electrodynamics \cite{Demianski1986}, and persists in the presence of a cosmological constant.

\noindent In the present work, Jacobson's argument is applied to black holes in the light of the running couplings, analysing the effect of this on the underlying physics as well as the implications associated with the specific form of gravitational coupling $G(r)$. Thus, for an effective Einstein-Hilbert action with cosmological constant, one has the effective Einstein field equations which read \cite{Koch:2016uso,Rincon:2017ypd}
\begin{align}\label{EOM}
G_{\mu \nu} + \Lambda(r)g_{\mu \nu} &= 8 \pi G(r) T_{\mu \nu}^{\text{effec}},
\end{align}
and the effective energy momentum tensor is defined according to
\begin{align}
8 \pi G(r) T_{\mu \nu}^{\text{effec}} &= 8 \pi G(r) T_{\mu \nu}^m -  \Delta t_{\mu \nu}.
\end{align}
Here, $T_{\mu \nu}^m$ is the matter energy momentum tensor (which for simplicity is taken to be zero) and the additional term $\Delta t_{\mu \nu}$ is related to the scale dependence of the gravitational running coupling $G(r)$ as shown in next section.

\section{Technique and Results}\label{Technique}

In spherical symmetry \ref{line_element} and according to Jacobson's idea \cite{Jacobson:2007tj}, the Null Energy Condition reads
\begin{align}\label{NEC}
T^{\text{effec}}_{\mu \nu} \ell^{\mu} \ell^{\nu} &= 0, \hspace{1cm} \therefore \hspace{1cm} R_{\mu \nu} \ell^{\mu} \ell^{\nu} = 0,
\end{align}
where, in three dimensional spacetime, $\ell^{\mu}$ is a null vector given by
\begin{align}
\ell^{\mu} \equiv \Bigl(\sqrt{g(r)},\ \sqrt{f(r)}, \ 0\Bigl), 
\end{align}
For arbitrary dimension, one always can define appropriate a null vector such as $\ell^{\mu} = (g^{1/2}, f^{1/2}, \bf{0})$, where $\bf{0}$ encoded the zero component for an arbitrary dimension of this null vector. Using the aforementioned $\ell^{\mu}$ combined with Eq. \ref{NEC}, it is ever possible to obtain a simple differential equation which is straightforward to solve.

\noindent Besides, note that in the Einstein-Hilbert truncation \cite{Lauscher:2002sq} the field equation of motion are consistent with Eq. \ref{EOM}. Thus, from \ref{NEC} one sees that 
\begin{align}
\bigg[R_{\mu \nu} - \bigg(\frac{1}{2}R - \Lambda(r)\bigg)g_{\mu \nu}\bigg] \ell^{\mu} \ell^{\nu} &= \bigg[8 \pi G(r) T_{\mu \nu}^m -  \Delta t_{\mu \nu}\bigg] \ell^{\mu} \ell^{\nu},
\end{align}
where the additional object $\Delta t_{\mu \nu}$ is defined as follows
\begin{align}
\Delta t_{\mu \nu} \equiv G_{k} \Big( g_{\mu \nu}\Box - \nabla_{\mu} \nabla_{\nu} \Big) G_{k}^{-1} .
\end{align}
Note that, by definition, a null vector satisfies that $g_{\mu \nu} \ell^{\mu} \ell^{\nu} = 0$, which allows getting the simple equation
\begin{align}\label{Null_tmunu}
R_{\mu \nu}\ell^{\mu}\ell^{\nu} &=0, \hspace{1cm} \therefore \hspace{1cm}  \Delta t_{\mu \nu}\ell^{\mu}\ell^{\nu} = 0,
\end{align}
which produces the following ordinary differential equation:
\begin{align}
2\left[\frac{dG(r)}{dr}\right]^2 - G(r)\frac{d^2G(r)}{dr^2} &= -\frac{1}{2}G(r)\frac{dG(r)}{dr}
\Bigg[\Bigl(f(r)\cdot g(r)\Bigl)^{-1}\frac{d}{dr}\Bigl(f(r) \cdot g(r)\Bigl)\Bigg],
\end{align}
thereby, one finds without even solving the set of equations
\ref{EOM} that 
\begin{align}\label{G_integral}
G(r) &= a\Bigg[\int_{r_0}^{r}\sqrt{f(r') \cdot g(r')} \ dr' \Bigg]^{-1},
\end{align}
independent of the actual form of $f(r)$ and $g(r)$. In addition,  $a$ is a real value which is taken such as one recovering $G(r) \rightarrow G_0$ in some limit.
Particular attention must be dedicated to case when the Schwarzschild ansatz is used. Under this assumption, the aforementioned differential equation is simplified to be
\begin{align}
2 \left[\frac{dG(r)}{dr}\right]^2 &= G(r) \frac{d^2G(r)}{dr^2},
\end{align}
which gives the following scale dependent gravitational coupling
\begin{align}\label{G_final}
G(r)= \frac{G_0}{1 + \epsilon r}.
\end{align}
The running coupling found in \cite{Koch:2016uso,Rincon:2017ypd,Rincon:2017goj,Koch:2015nva} corresponds to $f(r) \cdot g(r) = 1$, which is consistent with Eq. \ref{G_final}. Please, note that the relation $f(r) \cdot g(r) = 1$ is actually independent of the truncation and form of E.O.M.'s since by \ref{Null_tmunu} alone one finds the solution.
\section{Conclusion}\label{Conclusion}
   In the present work, the role of Null Energy Condition  \cite{Jacobson:2007tj} is analyzed in the context of running couplings. It is found that the NEC allows to obtain a justification of the Schwarzschild ansatz (by the virtue of \cite{Koch:2016uso}). By imposing non-generation of stress energy tensor due to scale dependent gravitational coupling \cite{Koch:2016uso,Rincon:2017ypd} it allows to determine the form of $G(r)$.

\section*{Acknowledgments}
The author A.R. was supported by the CONICYT-PCHA/Doctorado Nacional/2015-21151658
whereas the work of B.K. was supported by the Fondecyt 1161150.

\section*{References}
\bibliographystyle{iopart-num}
\bibliography{refsBTZ,myproyect}

\providecommand{\newblock}{}
\begin{thebibliography}{10}
\expandafter\ifx\csname url\endcsname\relax
  \def\url#1{{\tt #1}}\fi
\expandafter\ifx\csname urlprefix\endcsname\relax\def\urlprefix{URL }\fi
\providecommand{\eprint}[2][]{\url{#2}}

\bibitem{book:217893}
Rovelli C 2007 {\em Quantum gravity\/} Cambridge Monographs on Mathematical
  Physics (Cambridge University Press)

\bibitem{Polchinski:1998rq}
Polchinski J 2007 {\em {String theory. Vol. 1: An introduction to the bosonic
  string}\/} (Cambridge University Press)

\bibitem{Polchinski:1998rr}
Polchinski J 2007 {\em {String theory. Vol. 2: Superstring theory and
  beyond}\/} (Cambridge University Press)

\bibitem{Wheeler:1957mu}
Wheeler J~A 1957 {\em Annals Phys.\/} {\bf 2} 604--614

\bibitem{Deser:1976eh}
Deser S and Zumino B 1976 {\em Phys. Lett.\/} {\bf B62} 335

\bibitem{Rovelli:1997yv}
Rovelli C 1998 {\em Living Rev. Rel.\/} {\bf 1} 1 (\textit{Preprint}
  \eprint{gr-qc/9710008})

\bibitem{Bombelli:1987aa}
Bombelli L, Lee J, Meyer D and Sorkin R 1987 {\em Phys. Rev. Lett.\/} {\bf 59}
  521--524

\bibitem{Ashtekar:2004vs}
Ashtekar A 2005 {\em New J. Phys.\/} {\bf 7} 198 (\textit{Preprint}
  \eprint{gr-qc/0410054})

\bibitem{Sakharov:1967pk}
Sakharov A~D 1968 {\em Sov. Phys. Dokl.\/} {\bf 12} 1040--1041 [Gen. Rel.
  Grav.32,365(2000)]

\bibitem{Jacobson:1995ab}
Jacobson T 1995 {\em Phys. Rev. Lett.\/} {\bf 75} 1260--1263 (\textit{Preprint}
  \eprint{gr-qc/9504004})

\bibitem{Verlinde:2010hp}
Verlinde E~P 2011 {\em JHEP\/} {\bf 04} 029 (\textit{Preprint}
  \eprint{1001.0785})

\bibitem{Reuter:1996cp}
Reuter M 1998 {\em Phys. Rev.\/} {\bf D57} 971--985 (\textit{Preprint}
  \eprint{hep-th/9605030})

\bibitem{Litim:2003vp}
Litim D~F 2004 {\em Phys. Rev. Lett.\/} {\bf 92} 201301 (\textit{Preprint}
  \eprint{hep-th/0312114})

\bibitem{Horava:2009uw}
Horava P 2009 {\em Phys. Rev.\/} {\bf D79} 084008 (\textit{Preprint}
  \eprint{0901.3775})

\bibitem{Charmousis:2009tc}
Charmousis C, Niz G, Padilla A and Saffin P~M 2009 {\em JHEP\/} {\bf 08} 070
  (\textit{Preprint} \eprint{0905.2579})

\bibitem{Ashtekar:1981sf}
Ashtekar A 1981 {\em Phys. Rev. Lett.\/} {\bf 46} 573--576

\bibitem{Penrose:1986ca}
Penrose R and Rindler W 1988 {\em {SPINORS AND SPACE-TIME. VOL. 2: SPINOR AND
  TWISTOR METHODS IN SPACE-TIME GEOMETRY}\/} (Cambridge University Press) ISBN
  9780521347860, 9780511868429

\bibitem{Connes:1996gi}
Connes A 1996 {\em Commun. Math. Phys.\/} {\bf 182} 155--176 (\textit{Preprint}
  \eprint{hep-th/9603053})

\bibitem{Nicolini:2008aj}
Nicolini P 2009 {\em Int. J. Mod. Phys.\/} {\bf A24} 1229--1308
  (\textit{Preprint} \eprint{0807.1939})

\bibitem{Gambini:2004vz}
Gambini R and Pullin J 2005 {\em Phys. Rev. Lett.\/} {\bf 94} 101302
  (\textit{Preprint} \eprint{gr-qc/0409057})

\bibitem{Koch:2014joa}
Koch B, Rioseco P and Contreras C 2015 {\em Phys. Rev.\/} {\bf D91} 025009
  (\textit{Preprint} \eprint{1409.4443})

\bibitem{Koch:2010nn}
Koch B and Ramirez I 2011 {\em Class. Quant. Grav.\/} {\bf 28} 055008
  (\textit{Preprint} \eprint{1010.2799})

\bibitem{Koch:2016uso}
Koch B, Reyes I~A and Rinc\'on A 2016 {\em Class. Quant. Grav.\/} {\bf 33}
  225010 (\textit{Preprint} \eprint{1606.04123})

\bibitem{Contreras:2013hua}
Contreras C, Koch B and Rioseco P 2013 {\em Class. Quant. Grav.\/} {\bf 30}
  175009 (\textit{Preprint} \eprint{1303.3892})

\bibitem{Rincon:2017ypd}
Rinc\'on A, Koch B and Reyes I 2017 {\em J. Phys. Conf. Ser.\/} {\bf 831}
  012007 (\textit{Preprint} \eprint{1701.04531})

\bibitem{Jacobson:2007tj}
Jacobson T 2007 {\em Class. Quant. Grav.\/} {\bf 24} 5717--5719
  (\textit{Preprint} \eprint{0707.3222})

\bibitem{Demianski1986}
Demianski M 1986 {\em Foundations of Physics\/} {\bf 16} 187--190 ISSN
  1572-9516 \urlprefix\url{http://dx.doi.org/10.1007/BF01889380}

\bibitem{Lauscher:2002sq}
Lauscher O and Reuter M 2002 {\em Phys. Rev.\/} {\bf D66} 025026
  (\textit{Preprint} \eprint{hep-th/0205062})

\bibitem{Rincon:2017goj}
Rinc\'on A, Contreras E, Bargue\~no P, Koch B, Panotopoulos G and
  Hern\'andez-Arboleda A 2017  (\textit{Preprint} \eprint{1704.04845})

\bibitem{Koch:2015nva}
Koch B and Rioseco P 2016 {\em Class. Quant. Grav.\/} {\bf 33} 035002
  (\textit{Preprint} \eprint{1501.00904})

\end{thebibliography}

\end{document}